# Simulation Based Evaluation and Optimization for PEV Charging Stations Deployment in Transportation Networks

Zhenghe Zhong[1, †], Xinran Zhang[1, †], Daihan Zhang[1, †], Huimiao Chen[1, 2, *], Chuning Gao[1]

[1]Sparkzone Institute, Beijing, China
[2]Department of Electrical Engineering, Tsinghua University, Beijing, China
[†]Authors with equal contribution to this work
[*]Corresponding author (email: hchen@sparkzone.org)

*Abstract*— **Plug-in electric vehicles (PEVs) are receiving global attention. However, large-scale PEVs bring challenge to charging station deployment. This paper provides a deployment evaluation method and a planning method for PEVs charging stations based on simulation and genetic algorithm (GA). In this research, reasonable user behaviors changes i.e., detouring for recharging, is included in the method for more practical application. A detailed logic for describing the PEV users' detour behaviors is designed in this paper for the charging station deployment evaluation and further influence the later charging station planning strategy formulation. Intuitively, by taking the behaviors change of PEV users after a charging station deployment is given into account, our evaluation method is the more reasonable. Actually a series of control experiments should be carried out to illustrate the different insights from our method. However, in this paper, due to the limited space, we just provide the result of our method in a relatively simple case with some analyses. Different insights from our method may work as useful suggestion for charging station plan in a city area.**

*Keywords*—**Plug-in electric vehicle, charging station, user behaviors, evaluation, optimization, genetic algorithm.**

## I. Introduction

Topics pertinent to how the escalating environmental issues can be relieved have become ubiquitous in contemporary society, among which the study on popularizing plug-in electric vehicles (PEVs) has gained worldwide attention. With the higher awareness of the limited fossil fuels, researchers have found numerous approaches to reducing petroleum dependency, in which one effective solution is the transition from gasoline-powered vehicles to PEVs. Among several advantages of PEVs, the most prominent is low energy consumption and zero carbon emission. As compared with petroleum, electricity can come from renewable energy, such as wind and solar energy. With the rising popularity of PEVs, there comes a demand for a well-planned charging infrastructure network, a challenging task for researchers and engineers due to the peculiarity of PEVs. Recharging PEVs, unlike refueling gasoline-powered vehicles, takes significantly longer time and leads to a relatively shorter driving range. The anomaly of PEVs causes difficulties in the implementation of charging stations into the current infrastructures.

As PEVs are rapidly introduced to the market, particularly encouraged by the government policies, the charging stations must be planned and constructed in time, or else PEVs would not gain the expected market share. Such issue includes the demand for more accessible and well-placed charging stations. Without a comprehensive planning of charging facilities, the green experience would not be so desirable. Time-consuming queues and risks of not reaching destinations cause gasoline-powered vehicle users to hesitate towards going green, which becomes a stumbling rock on the road to sustainable energies. Hence, a thorough plan for PEV charging facilities is vital to the goal of an eco-friendly society.

Researchers have formulated a multitude of mathematical models for optimizing charging station deployments [1]-[9], as a high standard of efficiency must be achieved in city planning. The researches can be categorized into two different focuses: 1) based on power systems (e.g., strategies on individual charging stations to prevent overload on the power grid optimizes the pressure on power systems); 2) emphasis of transportation networks (e.g., some researches focusing on the transportation networks present comprehensive planning of charging stations in road networks to improve PEVs' user experience, thereby popularizing the greener travel mode).

In the power systems approach, the emphasis of individual charging stations addresses concerns on power loads. Authors of [1] propose a multi-objective EV charging station planning method, which ensures charging services while protecting the electrical systems. Then, a data-envelopment analysis method and a cross-entropy method are utilized to come to a complete charging station deployment. Research was done in [2] based on charging characteristics and trip characteristics of PEVs, where the authors calculate each station's power demand, and associates each station with the general power systems, then with such information of power demand, through a particle swarm optimization (PSO) and a weighted Voronoi diagram, generates a reasonably located and loaded planning of charging stations. As a goal to not only reduce the investment and operation cost of the system concerned but also promote the popularization of environmentally friendly PEVs, reference [3] utilizes a multi-objective collaborative planning strategy, then with integrations of the user equilibrium based traffic assignment model and a decomposition based multi-objective evolutionary algorithm, generates a suitable planning of PEV charging stations.

On the other hand, some researchers focus on the implementation of charging stations in road networks, concerning about waiting time, travel range, and trip efficiency. Integrating the optimization of queuing, reference [4] introduces a spatial and temporal modelling method to describe PEV charging demand, through a fluid dynamic traffic model and the M/M/s queueing theory, finally reaching a circumspect conclusion of charging station setting since each may vary by space and time. Considering the drawbacks of PEVs, including short ranges and underdeveloped supporting infrastructures, the authors of [5] seek solutions through a developed optimization model. Reference [6] approaches the problems of user experience with the grid partition method followed with the genetic algorithm (GA), formulates reasonable charging station plans in a designated area. In [7], based on an analytic hierarchy process, candidate sites are weighted, considering variables such as the distance between stations, installation cost, the operation costs, etc. The authors use a PSO algorithm, reaching a feasible result of charging stations planning. While most researches strictly focus on pure electric vehicles, some others have done studies on plug-in hybrid electric vehicles (PHEVs) (greenhouse gas is also reduced and the vehicles still maintain a considerable range). With PHEVs as the subject, the authors of [8] implement a model of time-series simulation to derive how each charging station ought to be configured. With a large-scale trajectory data of 11,880 taxis in Beijing as a case study, reference [9] reaches its goal to demonstrate that travel patterns mined from big-data can inform public charging infrastructure development. Besides proving a point, the research also draws useful information, for example, parking hotspots are suitable charging station candidates as it indicates high charging demand.

Many traits of PEVs have been considered during the process of research, as a slight discrepancy in electric vehicles can cause significant issues in terms of range, efficiency, and safety. The previous researches form a comprehensive solution to promoting PEVs, which propels the popularization of the green travel style. The power systems approach considers the stress on the power system caused by PEV charging, prevents risks on the power supply and optimizes energy consumption; while the transportation network approach focuses on queuing scenario and time consumption, provides superlative user experiences, simultaneously puts emphasis on charging station deployments, and opts the locations for the most user-friendly experiences. Although all the previous researches formulated commendable results, minor adjustments if implemented would provide results with higher veracity. The authors of [5] specify assumptions such as users have the option to swap batteries in the charging stations. However, battery electric vehicles no longer dominate the electric vehicle market, as battery electric vehicles are less effective in protecting the environment, such assumption can no longer be valid in the contemporary setting.

In [8] and [9], as a prerequisite, the authors assume "travel behaviors of drivers remain unchanged after adopting PHEVs." For PHEVs, since the range is less of an issue compared to PEVs, the clause may be a comprehensive assumption. However, the range of PEVs plays a significant factor in driver behaviors, as charging both require detours and consume more time.

In this research, based on simulation and reasonable PEV user behavior assumptions, we fully integrate traveler behavior changes, as the users must convert to a new practice due to the shorter range and less accessible charging facilities of PEVs. A rating (evaluation) method is introduced, which formulates a comprehensive score for each charging station deployment plan. Then, we leveraged the generic algorithm to arrive at a possibly optimal charging facility deployment.

In the reminder of the paper, Section II presents the evaluation framework for PEV charging station deployment in transportation networks. Based on the evaluation method, Section III shows a GA to obtain the PEV charging station planning strategy. Case studies is given in Section IV and Section V concludes.

## II. EVALUATION FRAMEWORK FOR PEV CHARGING STATION DEPLOYMENT

For a given charging station deployment, we present a method to evaluate its efficiency. And in our method, objective is to have the least unsatisfied energy.

We use circles and straight lines to describe traffic networks and a road-node incidence matrix to describe the connection relationships between nodes and roads. Fig. 1 and the matrix in Table I serves as a simple illustration.

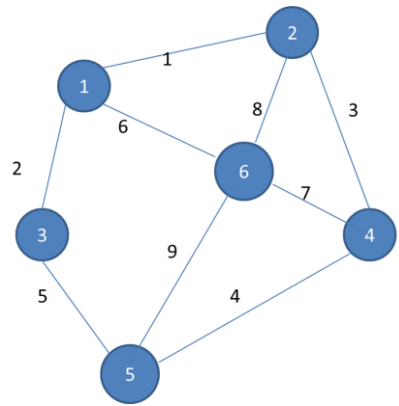

Fig. 1. A simple transportation network.

To simulate a real traffic network, we set clusters of nodes as origins and destinations, since in real life, there are residential areas, commercial areas, and industrial areas, which are usually separated. We define each trip data sample includes a origin, a destination, and the nodes and the roads between them. For example, in Fig. 1, 1→4→6 can be used to denote such a trip, where node 1 is the origin and node 6 is the destination.

Then, with a given charging station deployment, we develop the following method to assess a PEV's uncaptured state of charge (SOC). Use $SOC_{ini}$ to denote the SOC of the vehicle at origin node.

TABLE I. ROAD-NODE INCIDENCE MATRIX.

| Route \ Node | 1 | 2 | 3 | 4 | 5 | 6 | 7 | 8 | 9 |
|---|---|---|---|---|---|---|---|---|---|
| 1 | 1 | 1 |   |   |   | 1 |   |   |   |
| 2 | 1 |   | 1 |   |   |   |   | 1 |   |
| 3 |   | 1 |   |   | 1 |   |   |   |   |
| 4 |   |   | 1 | 1 |   |   | 1 |   |   |
| 5 |   |   |   | 1 | 1 |   |   |   | 1 |
| 6 |   |   |   |   |   | 1 | 1 | 1 | 1 |

*Scenario One*

A PEV can travel from its origin to destination without running out of battery power, i.e., there is no charging demand for this PEV while driving. In this case, its unsatisfied SOC is referred as $SOC_1$, which has a value of 0.

$$SOC_1 = 0 \quad (1)$$

*Scenario Two*

If the PEV cannot finish the route, then it must take a detour or multiple detours to recharge. As a basic premise of the model, we set that the PEV must return to the original route after the detour, which means that every node in the original route must be passed by the PEV even with detour (actually some nodes can be set as must to be passed and other not). For example, if a PEV original route is 6→4→2 and its $SOC_{ini}$ is not enough to drive to node 4, then it can take a detour to charge at node 1 and then go back to node 4 and finish the original route. However, the PEV cannot go directly from node 1 to node 2 because node 4 must be passed.

For a vehicle with a known initial SOC and route, we first find the number of nodes n that the PEV needs to take detour at. In this case, the car can choose to take 1 to n detours, and we compare the required energy for each situation. n is the maximum detour the PEV can take, and taking n detour does not guarantee to reach destination. But the PEV cannot reach the destination if it takes less than n detour. To evaluate the required energy, we first assume the vehicle takes m detour ($1 \le m \le n$). Then if the PEV finishes the whole route, the unsatisfied SOC equals to the extra SOC caused by the m detours.

Assuming that all drivers are rational, $SOC_{ini}$ should be no less than the energy required to reach the next node, or the drivers would not choose to drive.

If the PEV takes m detours and still cannot reach the destination, the unsatisfied SOC equals to the extra SOC caused by the m detours $SOC_{detour}$ times a coefficient $\alpha$ and then plus $SOC_{rest}$ times a coefficient $\beta$, in which $SOC_{rest}$ denotes the SOC needed to finish the rest of the route. We will indicate the coefficients in the remaining sections.

We refer the unsatisfied SOC of a car taking m detours as $E_m$. Through the method we indicated above, we are able to evaluate the unsatisfied SOC for every situation: because the PEV cannot reach the destination by taking less than n detour, thus $E_m = \alpha\, SOC_{detour} + \beta\, SOC_{rest}$ ($1 \le m \le n-1$); if it reaches the destination aftrer taking n detour, then $E_n = SOC_{detour}$, else $E_n = \alpha\, SOC_{detour} + \beta\, SOC_{rest}$. Thus, for all the situations, we choose the one with the least unsatisfied SOC.

$$SOC_2 = \min(E_0, E_1, E_2, \ldots\ldots E_n) \quad (2)$$

Another important premise is that a PEV charges only once during one detour. Charging more than once is inefficient in regards to both time and energy. Is it also not likely for a PEV to charge more than once in one detour, taking into consideration of PEVs' driving range and the distance of a detour. Also, we assume that charging stations can only be built at nodes, and if a charging station is built between two nodes, we regard it as a new node.

To sum up, we use Fig. 2, a flow chart, as a general and clear illustration.

With our method of assessing unsatisfied SOC, we can compare the performance of different charging station deployment and then select the optimal one. In the next section, we use GA to get a suboptimal solution.

### III. GENETIC ALGORITHM FOR CHARGING STATION SITING OPTIMIZATION IN TRANSPORTATION NETWORKS

In this section, genetic algorithm is implemented and utilized to calculate an optimal plan for charging stations distribution in transportation network.

*Initialization*

The initial population is randomly generated by the pop function, where each individual is given a length and number of 1s. The size of each chromosome represents total station candidates and numbers of stations to be built. Consists of 0s and 1s, the chromosomes represent charging stations deployments in the designated area. The nodes of the chromosomes reflect the occupation of charging station candidates through a binary representation, where 0 represent no station built and 1 shows the presence of a charging station. Each time a chromosome is generated, an if statement ensures it holds the set numbers of 0s and 1s and fills the population to the set number.

*Calculate Fit Value*

In a loop of numerous iterations, each individual representation of charging station plans is graded by the evaluation method presented in the previous section. With calculated evaluation results, each individual chromosome is given a fit value that represents its practicability and possibility of survival in the process of selection.

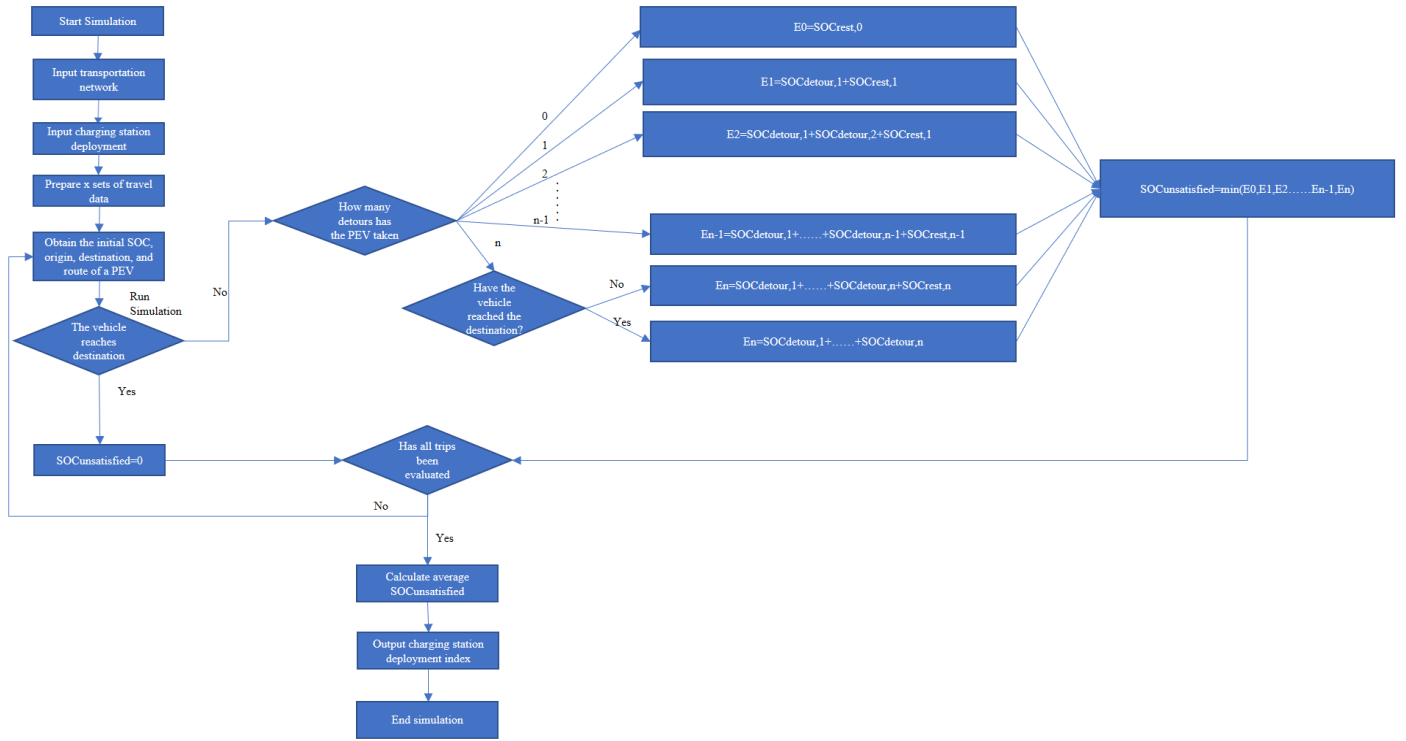

Fig. 2. The flow chart of all the evaluation method.

*Selection*

After fit values are assigned to individuals, each chromosome, correspondingly a distribution plan for stations, is given a probability calculated by its potential fitness in the whole population, with function 1. The probability is directly related to its fit value's ratio to the sum of all fit values, which forms a fitness roulette wheel that simulates natural selection in the natural world. The greater fit value a plan possesses, the better chance it would survive the iterations in the population. According to the value of roulette, individual chromosomes are drawn into the population until the initial population size is reached, maintaining the number of chromosomes in the population. However, duplications of chromosomes are possible in the population, as some fit in the evaluation method is much better than others.

$$p_{fitvalue} = fitvalue / totalfit \qquad (2)$$

*Crossover*

During the processes of genetic crossovers, a predetermined crossover probability restricts the occurrences of crossovers. When the randomized coefficient meets a standard for a crossover to happen, two adjacent individual chromosomes are scanned for a given length with the same number of deployed stations, to ensure that the crossover does not alter the number of 1s in each chromosome. Then, the sections are swapped to generate two new possible configurations. The results of the genetic crossover method overwrite the parent chromosomes in the population, for the purpose to maintain the population size.

*Mutation*

The mutation swaps two nodes in one parent chromosome when the mutation probability allows. The two nodes being swapped are ensured to be a 0 and a 1 through a process, thus the new chromosome is different while the number of 1s remains constant. The process consists of for loops and if statements to test whether the mutated nodes have a different status of occupation or not, that the mutation method might alter the number of stations built or else keep the original plan unchanged. The results of the mutation method overwrite the parent chromosome in the population, to keep the original numbers of chromosomes in the population.

IV. CASE STUDIES

In this section, the proposed evaluation method and GA based charging station planning method are simulated. The key of our method is the more reasonable evaluation method, so actually a series of control experiments should be carried out to illustrate the different insights from our method, which should be a long report. Here, we just provide a result of our method in a case with some simple analyses.

*A. Scenario and parameter settings*

We perform our simulation based on the transportation networks shown in Fig. 4. We define that there are three types of areas, i.e., residential areas, commercial areas and other areas, in the city transportation networks. We assume the cluster of dark green nodes as residential areas, the cluster of black nodes as commercial areas, and the cluster of red nodes as other areas, respectively. We suppose that all the nodes in the networks are the candidate nodes where charging stations can be built. The number of final charging station is set as 5.

OD trips are then generated for the given transportation networks. Note that, for simplification, we consider the trips within a time period, e.g., a day. Considering that the travel demands between commercial and residential areas are relatively large compared with those between other two areas, we make the trips between these two areas denser. The detailed settings of traffic trip data, of which the format is aforementioned, are skipped here due to limited space.

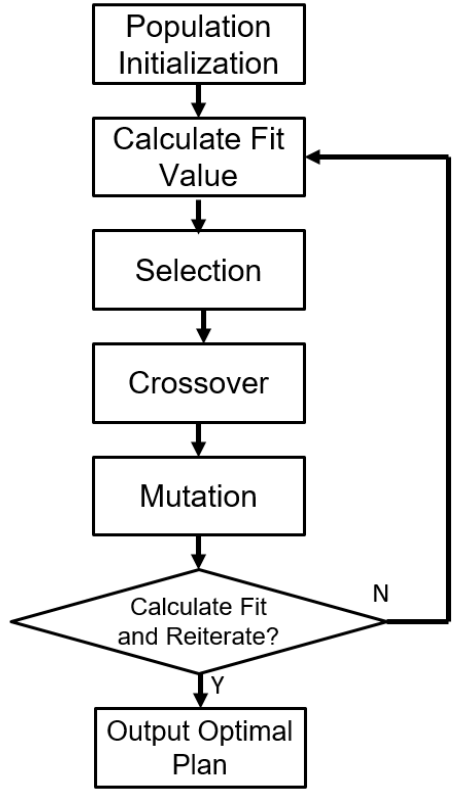

Fig. 3. GA flow chart.

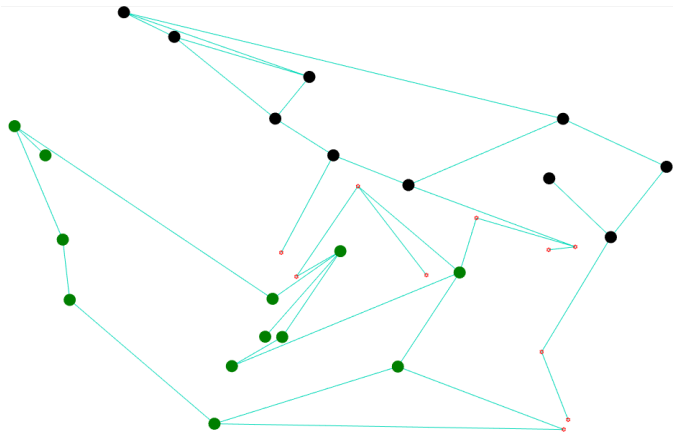

Fig. 4. Transportation networks for case studies.

## B. Simulation Results and Analysis

According to the above parameter settings and the model formulated in previous sections, we carry out the simulations using in Matlab R2016a. Fig 5 gives the result of selected five stations with yellow circles. Fig. 6 shows the fit value curve while using the GA.

We can observe that the station deployment has two traits: 1) distributed in different parts in the networks; 2) due to the demand between the residential and commercial areas are larger, the charging points aims to satisfy the demand of all the trips; 3) the charging station not always be located in nodes with high traffic flow, usually the nodes with large out-degree, because we consider the PEV users' behaviors change like detours in the method.

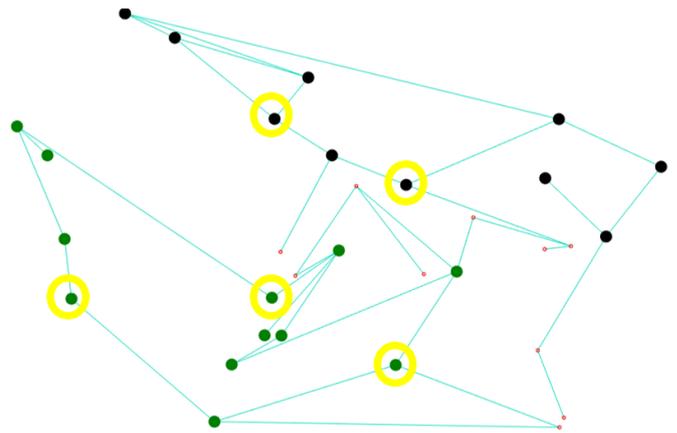

Fig. 5. Results of charging station choice.

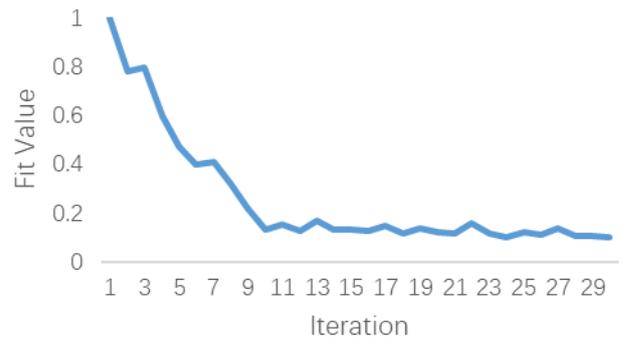

Fig. 6. Fit value curve.

## V. CONSLUSIONS AND FUTURE WORK

In this paper, we present an evaluation method for charging deployment based on simulation and reasonable PEV user behavior assumptions. Then, we leveraged the GA to obtain a possibly optimal charging facility deployment. In the future, more detailed case comparisons should be provided as well as the method refined.